\documentclass[aps,prd,preprint,groupedaddress,nofootinbib]{revtex4-1}
\usepackage{amsmath}
\usepackage{slashed}
\usepackage{amsfonts}
\usepackage{amsmath}
\usepackage{booktabs}
\usepackage{siunitx}
\usepackage{multirow}
\usepackage{colortbl}
\usepackage{xcolor}
\usepackage{graphicx}
\usepackage[utf8]{inputenc}
\usepackage{enumitem}
\usepackage{booktabs}
\usepackage{siunitx}
\usepackage{multirow}
\usepackage[%
  colorlinks=true,
  urlcolor=blue,
  linkcolor=blue,
  citecolor=blue
  ]{hyperref}
\usepackage{cleveref}

\newcolumntype{C}{>{$}c<{$}}
\AtBeginDocument{
\heavyrulewidth=.08em
\lightrulewidth=.05em
\cmidrulewidth=.03em
\belowrulesep=.65ex
\belowbottomsep=0pt
\aboverulesep=.4ex
\abovetopsep=0pt
\cmidrulesep=\doublerulesep
\cmidrulekern=.5em
\defaultaddspace=.5em
}
\begin{document}


\title{Reanalysis of $\Omega_c \rightarrow \Xi_c^+ K^-$ decay in QCD}


\author{T.~M.~Aliev}
\email[]{taliev@metu.edu.tr}
\altaffiliation{Permanent Address: Institute of Physics, Baku, Azerbaijan}
\affiliation{Department of Physics, Middle East Technical University, 06800, Ankara, Turkey}

\author{S.~Bilmis}
\email[]{sbilmis@metu.edu.tr}
\affiliation{Department of Physics, Middle East Technical University, 06800, Ankara, Turkey}

\author{M.~Savci}
\email[]{savci@metu.edu.tr}
\affiliation{Department of Physics, Middle East Technical University, 06800, Ankara, Turkey}


\date{\today}

\begin{abstract}
  The strong coupling constants of newly observed $\Omega_c^0$ baryons with spin $J = \frac{1}{2}$ and $J = \frac{3}{2}$ decaying into $\Xi_c^{+} K^{-}$ are estimated within light cone QCD sum rules. The calculations are performed within two different scenarios on quantum numbers of $\Omega_c$ baryons: a)  all newly observed $\Omega_c$ baryons are negative parity baryons, i.e., the $\Omega_c(3000)$, $\Omega_c(3050)$, $\Omega_c(3066)$, and $\Omega_c(3090)$ have quantum numbers $J^P = \frac{1}{2}^-$ and $J^P = \frac{3}{2}^-$ states respectively; b) the states $\Omega_c(3000)$ and $\Omega_c(3050)$ have quantum numbers $J^P = \frac{1}{2}^-$ and $J^P = \frac{1}{2}^+$
  , while the states $\Omega_c(3066)$ and $\Omega_c(3090)$ have the quantum numbers $J^P = \frac{3}{2}^-$ and $J^P = \frac{3}{2}^+$, respectively. By using the obtained results on the coupling constants, we calculate the decay widths of the corresponding decays.  The results on decay widths are compared with the experimental data of LHC Collaboration. We found out that the predictions on decay widths within these scenarios are considerably different from the experimental data, i.e., both considered scenarios are ruled out.
\end{abstract}


\maketitle

\section{Introduction}
\label{sec:intro}
Lately, in the invariant mass spectrum of $\Xi_c^{+} K^{-}$, very narrow excited $\Omega_c$ states ($\Omega_c(3000)$, $\Omega_c(3050)$, $\Omega_c(3066)$, $\Omega_c(3090)$, and $\Omega_c(3119)$) have been observed at LHCb~\cite{Aaij:2017nav}. Quantum numbers of these newly observed states have not been determined in the experiments yet. Hence, various possibilities about the quantum numbers of these states have been speculated in recent works. In~\cite{Agaev:2017jyt}, the states $\Omega_c(3050)$ and $\Omega_c(3090)$ are assigned as radial excitation of ground state $\Omega_c(3000)$ and
$\Omega^*(3066)$ baryons with the $J^P= \frac{1}{2}^{+}$ and $\frac{3}{2}^{+}$, respectively. On the other hand, in \cite{Karliner:2017kfm,Wang:2017vnc,Padmanath:2017lng,Wang:2017zjw,Aliev:2017led} these new states are assumed as the $P$-wave states with $J^P = \frac{1}{2}^{-}, \frac{1}{2}^{-}, \frac{3}{2}^{-}, \frac{3}{2}^{-}$ and $\frac{5}{2}^{-}$ respectively. Moreover the new states are assumed as pentaquarks in~\cite{Yang:2017rpg}. Similar quantum numbers of these new states are assigned in~\cite{Wang:2017hej}. Analysis of these states is also studied with lattice QCD, and the results indicated that most probably these states have $J^P = \frac{1}{2}^{-},~\frac{3}{2}^{-},~\frac{5}{2}^{-}$ quantum numbers~\cite{Padmanath:2017lng}. Another set of quantum number assignments, namely $\frac{3}{2}^{-}$, $\frac{3}{2}^{-}$, $\frac{5}{2}^{-}$, and $\frac{3}{2}^{+}$ is given in~\cite{Karliner:2017kfm}. In~\cite{Wang:2017xam}, it is obtained that the prediction on mass supports assigning $\Omega_c(3000)$ as $J^P = \frac{1}{2}^{-}$, $\Omega_c(3090)$ as $J^P = \frac{3}{2}^{-}$ or the $2S$ state with $J^P = \frac{1}{2}^+$, and $\Omega_c(3119)$ as $J^P = \frac{3}{2}^+$.

In this work, we estimate the strong coupling constants of $\Omega_c^0 \rightarrow \Xi_c^{+} K^{-}$ in the framework of light-cone QCD sum rules.
In our calculations, two different possibilities on quantum numbers of $\Omega_c$ baryons are explored:
\begin{enumerate}[label=\alph*)]
\item All newly observed $\Omega_c$ states have negative parities. More precisely, $\Omega_c(3000)$, $\Omega_c(3050)$ have quantum numbers $J^P = \frac{1}{2}^{-}$, $\Omega_c(3066)$, $\Omega_c(3090)$ have $\frac{3}{2}^{-}$, and $\Omega_c(3119)$ has quantum numbers $\frac{5}{2}^{-}$.
\item Part of newly observed $\Omega_c$ baryons have negative parity, and another part represents as a radial excitations of ground state baryons, i.e., $\Omega_c(3000)$ has $J^P = \frac{1}{2}^{-}$, $\Omega_c(3050)$ has $J^P = \frac{1}{2}^+$;  $\Omega_c(3066)$ and $\Omega_c(3090)$ states have quantum numbers $J^P = \frac{3}{2}^{-}$ and $\frac{3}{2}^{+}$, respectively.
\end{enumerate}

Note that the strong coupling constants of $\Omega_c \rightarrow \Xi_c^{+} K^{-}$ decays within the same framework are studied in~\cite{Agaev:2017lip}, and in chiral quark model~\cite{Ball:2006wn} respectively. However, the analysis performed in \cite{Agaev:2017lip} is incomplete. First of all, the contribution of negative parity $\Xi_c$ baryons is neglected entirely. Second, in our opinion the numerical analysis presented in~\cite{Agaev:2017lip} is inconsistent.

The article is organized as follows. In section~\ref{sec:2} the light cone sum rules for the coupling constants of $\Omega_c \rightarrow \Xi_c^{+} K^{-}$ decays are derived. Section~\ref{sec:numeric} is devoted to the analysis of the sum rules obtained in the previous section. In this section, we also estimate the widths of corresponding decays, and comparison with the experimental data is presented.
\section{Light cone sum rules for the strong coupling constants of $\Omega_c \rightarrow \Xi_c^{+} K^-$  transitions} 
\label{sec:2}
For the calculation of the strong coupling constants of $\Omega_c \rightarrow \Xi_c^{+} K^-$ transitions we consider the following two correlation functions in both pictures,
\begin{equation}
  \label{eq:1}
  \Pi = i \int d^4x e^{ipx} \langle K(q) | \eta_{\Xi_c}(x) \bar{\eta}_{\Omega_c}(0) | 0 \rangle,
\end{equation}
and
\begin{equation}
  \label{eq:1a}
  \Pi^\mu = i \int d^4x e^{ipx} \langle K(q) | \eta_{\Xi_c}(x) \bar{\eta}_{\Omega^{*}_{c}}^\mu(0) | 0 \rangle,
\end{equation}
where $\eta_{\Xi_c}~(\eta_{\Omega_c})$ is the interpolating current of $\Xi_c~(\Omega_c)$ baryon and $\eta_{\Omega^{*}_{c}}^\mu$ is the interpolating current of $ J^P = \frac{3}{2}$ $\Omega_c^*$ baryon:
\begin{equation}
  \label{eq:2}
  \begin{split}
    \eta_{\Xi_c}  = \frac{1}{\sqrt{6}} \epsilon^{abc} \bigg\{ & 2(u^{a^T}Cs^b) \gamma_5 c^c + 2\beta (u^{a^T} C \gamma_5 s^b)c^c + (u^{a^T}C c^b) \gamma_5 s^c \\
    & +\beta (u^{a^T}C \gamma_5 c^b) s^c + (c^{a^T}Cs^b) \gamma_5 u^c + \beta (c^{a^T} C \gamma_5 s^b) u^c \bigg\}
  \end{split}
\end{equation}

\begin{equation}
  \label{eq:3}
  \eta_{{\Omega_c}} = \eta_{\Xi_c} ( u \rightarrow s) ~~~~~~~~~~~~~~~~~~~~~~~~~~~~~~~~~~~~~~~~~~~~~~~~~~~~~~~~~~~~~~~~~~
\end{equation}

\begin{equation}
  \label{eq:4}
  \eta_{\Omega_c^{*}}^\mu = \frac{1}{\sqrt{3}} \epsilon^{abc} \bigg\{ (s^{a^T} C \gamma^\mu s^b) c^c + (s^{a^T} C \gamma^\mu c^b) s^c + (c^{a^T} C \gamma^\mu s^b) s^c \bigg\} ~~~~~~~~~
\end{equation}
where $a,~b$ and $c$ are color indices, $C$ is the charge conjugation operator and $\beta$ is arbitrary parameter.

We calculate $\Pi$ ($\Pi_\mu$) employing the light cone QCD sum rules (LCSR). According to the sum rules method approach, the correlation functions in Eqs.~\eqref{eq:1} and \eqref{eq:1a} can be calculated in two different ways:
\begin{itemize}
\item In terms of hadron parameters,
  \item In terms of quark-gluons in the deep Euclidean domain. 
  \end{itemize}
 These two representations are then equated by using the dispersion relation, and we get the desired sum rules for corresponding strong coupling constant. The hadronic representations of the correlation functions can be obtained by saturating Eqs.~\eqref{eq:2} and \eqref{eq:3} with corresponding baryons.

  Here we would like to note that the currents $\eta_{\Xi_c}$, $\eta_{\Omega_c}$, and $\eta_{\Omega_c^{*}}$ interact with both positive and negative parity baryons. Using this fact for the correlation functions from hadronic part we get
  \begin{equation}
    \label{eq:5}
      \Pi^{(\text{had})} = \sum_{\substack{i= +,- \\ j= +,-}} \frac{\langle 0 | \eta_{\Xi_c} | \Xi_c^j(p) \rangle \langle K(q) \Xi_c^{j}(p)| \Omega_c^{i}(p+q) \rangle\langle \Omega_c^{i}(p+q) | \bar{\eta}_{\Omega_c} |0 \rangle}{(p^2 - m_{\Xi_{c_j}}^2) \big( (p+q)^2 - m_{\Omega_{c}(i)}^2\big)},
  \end{equation}
  \begin{equation}
    \label{eq:6}
          \Pi^{\mu^{(\text{had})}} = \sum_{\substack{i= +,- \\ j= +,-}} \frac{\langle 0 | \eta_{\Xi_c} | \Xi_c^j(p) \rangle \langle K(q) \Xi_c^{j}(p)| \Omega_{c^*}^{i}(p+q) \rangle\langle \Omega_{c^*}^{i}(p+q) | \bar{\eta}_{\Omega_{c}}^\mu | 0 \rangle}{(p^2 - m_{\Xi_{c_j}}^2) \big( (p+q)^2 - m_{\Omega^{*}_{c}(i)}^2\big)}.
  \end{equation}
  The matrix elements in Eqs.~\eqref{eq:5} and \eqref{eq:6} are determined as
  \begin{equation}
    \label{eq:32}
    \begin{split}
      \langle 0 | \eta_{B} | B^{(+)}(p) \rangle = \lambda_{+} u(p), \\
      \langle 0 | \eta_{B} | B^{(-)}(p) \rangle = \lambda_{-} u(p),
    \end{split}
  \end{equation}

  \begin{equation}
    \label{eq:33}
    \begin{split}
        \langle K(q) B(p) | B(p+q) \rangle =  g \bar{u}(p) i \Gamma u(p+q), 
    \end{split}
  \end{equation}
  \begin{equation}
    \label{eq:35}
    \begin{split}
           \langle K(q) B(p) | B^{*}(p+q) \rangle = g^* \bar{u}(p) \Gamma^\prime u_\mu(p+q) q^\mu, 
    \end{split}
  \end{equation}
  where,
  \begin{equation}
    \label{eq:34}
    \Gamma = 
    \begin{cases}
      \gamma_5 & \text{for}~ -(+) \rightarrow -(+) \\
       1 & \text{for}~ -(+) \rightarrow +(-) ~~~\text{transitions},
    \end{cases}
  \end{equation}

  \begin{equation}
    \label{eq:36}
\Gamma^\prime =
\begin{cases}
1 & \text{for}~ -(+) \rightarrow -(+)\\
\gamma_5 & \text{for}~ -(+) \rightarrow +(-)~~~\text{transitions},
\end{cases}
  \end{equation}
  and $g$ is strong coupling constant of the corresponding decay, $\lambda_{B^{(i)}}$ are the residues of the corresponding baryons, $u_\mu$ is the Rarita-Schwinger spinors. Here the sign $+(-)$ corresponds to positive (negative) parity baryon. In further discussions, we will denote the mass and residues of ground and excited states of $\Omega_c(\Omega_c^*)$ baryons as: $m_0, \lambda_0 (m_0^*, \lambda_0^*)$, $m_1, \lambda_1 (m_1^*, \lambda_1^*)$, $m_2, \lambda_2 (m_2^*, \lambda_2^*)$ for scenario a); and for scenario b), the same notation is used as in previous case by just replacing $m_2, \lambda_2 (m_2^*, \lambda_2^*)$ to $m_3, \lambda_3 (m_3^*, \lambda_3^*)$. Moreover, the mass and residues of $\Xi_c$ baryons are denoted as $m^\prime_0$, $\lambda_0^\prime$ and $m^\prime_1, \lambda_1^\prime$. Using the matrix elements defined in \Cref{eq:32,eq:33,eq:35} for the correlation functions given in \Cref{eq:1,eq:1a} we get (for case a):


 \begin{equation}
    \label{eq:26}
    \begin{split}
      \Pi &= i A_1 (\slashed{p} + m_{0}^{\prime})  \gamma_5 (\slashed{p} + \slashed{q} + m_{0}) \\
      &+ i A_2 (\slashed{p} + m_{0}^{\prime})(\slashed{p} + \slashed{q} + m_{1})(-\gamma_5) \\
      &+ i A_3 \gamma_5 (\slashed{p} + m_{1}^{\prime})  (\slashed{p} + \slashed{q} + m_{0}) \\
      &+i A_4 \gamma_5 (\slashed{p} + m_{1}^{\prime}) \gamma_5 (\slashed{p} + \slashed{q} + m_{1})(-\gamma_5) \\
      &+i A_5(\slashed{p} + m_{0}^{\prime}) (\slashed{p} + \slashed{q} + m_{2})(-\gamma_5) \\
      &+i A_6 \gamma_5 (\slashed{p} + m_{1}^{\prime}) \gamma_5 (\slashed{p} + \slashed{q} + m_{2})(-\gamma_5)
    \end{split}
  \end{equation}

  \begin{equation}
    \label{eq:27}
    \begin{split}
      \Pi^\mu &= +A_1^{*} (\slashed{p} + m_{0}^{\prime}) (\slashed{p} + \slashed{q} + m_{0}^*) (-q^\mu) \\
              &+ A_2^{*} (\slashed{p} + m_{0}^{\prime}) \gamma_5 (\slashed{p} + \slashed{q} + m_{1}^*)(-\gamma_5)(-q^\mu)\\
              &+ A_3^{*} \gamma_5 (\slashed{p} + m_{1}^{\prime}) \gamma_5  (\slashed{p} + \slashed{q} + m_{0}^*) (-q^\mu) \\
              &+ A_4^{*} \gamma_5 (\slashed{p} + m_{1}^{\prime}) (\slashed{p} + \slashed{q} + m_{1}^*)(-\gamma_5) (-q^\mu) \\
               &+ A_5^{*}  (\slashed{p} + m_{0}^{\prime}) \gamma_5  (\slashed{p} + \slashed{q} + m_{2}^*) (-\gamma_5) (-q^\mu) \\
              &+ A_6^{*} \gamma_5 (\slashed{p} + m_{1}^{\prime}) ( \slashed{p} + \slashed{q} + m_{2}^*)(-\gamma_5) (-q^\mu) + \text{other structures},
    \end{split}
  \end{equation}
  where

  \begin{equation}
    \label{eq:13}
    \begin{split}
      A_1^{(*)} =& \frac{\lambda_{0}^{(*)} \lambda_{0}^{\prime}  g_1^{(*)}}{ ({m_{0}^{\prime}}^2 -  p^2 ) \big({m_{0}^{(*)}}^2 -(p+q)^2\big)}  \\
      A_2^{(*)} =& A_1 \big( m_0^{(*)} \rightarrow m_{1}^{(*)}, \hspace{4mm}  g_{1}^{(*)} \rightarrow g_{2}^{(*)}, \hspace{4mm}  \lambda_{0}^{(*)} \rightarrow \lambda_{1}^{(*)}  \big)\\
      A_3^{(*)} =& A_1 \big( m_0^{\prime} \rightarrow m_{1}^{\prime}, \hspace{4mm} g_{1}^{(*)} \rightarrow g_{3}^{(*)}, \hspace{4mm}  \lambda_{0}^{\prime} \rightarrow \lambda_{1}^{\prime} \big)\\
      A_4^{(*)} =& A_3 \big( m_0^{(*)} \rightarrow m_{1}^{(*)},  \hspace{4mm} g_{3}^{(*)} \rightarrow g_{4}^{(*)}, \hspace{4mm}  \lambda_{0}^{(*)} \rightarrow \lambda_{1}^{(*)} \big)\\
      A_5^{(*)} =& A_2 \big( m_{1}^{(*)} \rightarrow m_{2}^{(*)}, \hspace{4mm} g_{2}^{(*)} \rightarrow g_{5}^{(*)}, \hspace{4mm}  \lambda_{1}^{(*)} \rightarrow \lambda_{2}^{(*)} \big)\\
      A_6^{(*)} =& A_5 \big( m_0^{\prime} \rightarrow m_{1}^{\prime}, \hspace{4mm} g_{5}^{(*)} \rightarrow g_{6}^{(*)}, \hspace{4mm}  \lambda_{0}^{\prime} \rightarrow \lambda_{1}^{\prime} \big)
    \end{split}
  \end{equation}
  The result for the scenario b) can be obtained from Eqs.~\eqref{eq:26} and \eqref{eq:27} by following replacements:
  \begin{equation}
    \label{eq:29}
    \begin{split}
      A_5 (\slashed{p} + m_{0}^{\prime}) (\slashed{p} + \slashed{q} + m_{2})(-\gamma_5) & \rightarrow \tilde{A_5} (\slashed{p} + m_{0}^{\prime}) i\gamma_5 (\slashed{p} + \slashed{q} + m_{3}) \\
      A_6 \gamma_5 (\slashed{p} + m_{1}^{\prime}) \gamma_5 (\slashed{p} + \slashed{q} + m_{2}) (-\gamma_5) & \rightarrow \tilde{A_6} \gamma_5 (\slashed{p} + m_{1}^{\prime}) (\slashed{p} + \slashed{q} + m_{3}) \\
      A_5^* (\slashed{p} + m_{0}^{\prime}) \gamma_5 q^\alpha (\slashed{p} + \slashed{q} + m_{2}^{*})(-\gamma_5)(-g_{\mu \alpha}) & \rightarrow \tilde{A_5} (\slashed{p} + m_{0}^{\prime}) q^\alpha (\slashed{p} + \slashed{q} + m_{3}^{*}) (-g_{\mu \alpha}) \\
      A_6^* \gamma_5 (\slashed{p} + m_{1}^{\prime}) q^\alpha (\slashed{p} + \slashed{q} + m_{2}^{*})(-g_{\mu \alpha})(-\gamma_5) &\rightarrow \tilde{A_6} \gamma_5 (\slashed{p} + m_{1}^{\prime}) q^\alpha \gamma_5 (\slashed{p} + \slashed{q} + m_{3}^{*}) (-g_{\mu \alpha}).
    \end{split}
  \end{equation}

  Note that to derive Eq.~\eqref{eq:27}, we used the following formula for performing summation over spins of Rarita-Schwinger spinors
  \begin{equation}
    \label{eq:16}
    \sum u_\alpha(p) \bar{u}_\beta(p) = -(\slashed{p} + m) \bigg( g_{\alpha\beta}- \frac{\gamma_\alpha\gamma_\beta}{3} + \frac{2p_\alpha p_\beta}{3m^2} + \frac{p_\alpha \gamma_\beta - p_\beta \gamma_\alpha}{3m} \bigg)
  \end{equation}
  and in principle one can obtain the expression for the hadronic part of the correlation function. At this stage two problems arise. One of them is dictated by the fact that the current $\eta_\mu$ interacts not only with spin 3/2 but also 1/2 states. The matrix element of the current $\eta_\mu$ with spin $1/2$ state is defined as
  \begin{equation}
    \label{eq:17}
    \langle 0 | \eta_\mu | 1/2 \rangle = A \bigg( \gamma_\mu - \frac{4}{m}p_\mu \bigg) u(p),
  \end{equation}
  i.e., the terms in the RHS of Eq.\eqref{eq:16} $\sim \gamma_\mu$ and the right end $(p+q)_\mu$ contain contributions from $1/2$ states, which should be removed. The second problem is related to the fact that not all structures appearing in Eqs.~\ref{eq:27} are independent. In order to cure both these problems we need ordering procedure of Dirac matrices. In present work, we use ordering of Dirac matrices as $\slashed{p} \slashed{q} \gamma_\mu$. Under this ordering, only the term $\sim g_{\mu \alpha}$ contains contributions solely from spin $3/2$ states. For this reason, we will retain only $~g_{\mu \alpha}$ terms in the RHS of Eq.\eqref{eq:27}. 
  
  In order to find sum rules for the strong coupling constants of $\Omega_c \rightarrow \Xi_c^+ K^-$ transitions we need to calculate the $\Pi$ and $\Pi_\mu$ from QCD side in the deep Euclidean region, $p^2 \rightarrow - \infty$, $(p+q)^2 \rightarrow -\infty$. The correlation from QCD side can be calculated by using the operator product expansion.

  Now let demonstrate steps of calculation of the correlation function from QCD side. As an example let consider one term of correlation $\Pi_\mu$, i.e. consider

  \begin{equation}
    \label{eq:10}
    \Pi_\mu \sim \epsilon^{abc} \epsilon^{a_1 b_1 c_1} \int d^4 x e^{ipx} \langle K(q) | \frac{1}{\sqrt{6}} 2 (u^{a T} C \gamma^5 s^{b_1}) c^{c_1}(x) \big[ \bar{c}_{(0)}^{c_1} (\bar{s}^{b_1}(0) \gamma_\mu C \bar{s}^{a_1 T}) | 0 \rangle \big]
  \end{equation}

  By using Wick's theorem, this term can be written as;
  \begin{equation}
    \label{eq:11}
    \begin{split}
      \Pi_\mu \sim& \epsilon^{abc} \epsilon^{a_1 b_1 c_1} \int d^4 x e^{ipx} \langle K(q) | \bigg\{ \bar{s}^{a_1}_{1} (\gamma_\mu C)^T S_s^{b b_1 T}(x) C^T u^{a_1} (x) \gamma_5 S_c^{c c_1} (x)  \\
      &~~~~~~~~~~~~~~~~~~~~~~~~~~~~~~~~~~ - (\bar{s}^{b_1}(0) \gamma_\mu C S_s^{b a_1 T} C^T u^a ) \gamma_5 S_c^{c c_1}(x) \bigg\} | 0 \rangle
  \end{split}
  \end{equation}
  From this formula, it follows that to obtain the correlation function(s) from QCD side, first of all we need the expressions of light and heavy quark propagators. The expressions of the light quark propagator in the presence of gluonic and electromagnetic background fields are derived in \cite{Balitsky:1987bk}:
  \begin{equation}
    \label{eq:20}
    \begin{split}
      S(x)  = \frac{i \slashed{x}}{2 \pi^2 x^4} -\frac{m_q}{4 \pi x^2} - \frac{i g_s}{16 \pi^2} \int & du \bigg\{ \frac{\bar{u} \slashed{x} \sigma_{\alpha \beta} + u \sigma_{\alpha \beta} \slashed{x}}{x^2} [g_s G^{\alpha \beta}(ux) + e_q F^{\alpha \beta}] \\
      &-\frac{i m_q}{2}[g_s G_{\mu\nu} \sigma^{\mu \nu} + e q F_{\mu \nu} \sigma_{\mu \nu}] (ln \frac{-x^2 \Lambda^2}{4} + 2 \gamma_E) \bigg\}.      
    \end{split}
  \end{equation}

  The heavy quark propagator is given as,
  \begin{equation}
    \label{eq:21}
    \begin{split}
      S_Q = \int  \frac{d^4 k}{(2\pi)^4} e^{-ikx} \frac{i(\slashed{k}+m_Q)}{k^2-m_Q^2} - ig_s \int \frac{d^4k}{(2 \pi)^4 i} \int_0^{1}du & \big[ \frac{\slashed{k}+m_Q}{2(m_Q^2 - k^2)^2} G^{\mu \nu}(ux) \sigma_{\mu \nu} \\
      &+ \frac{i x_\mu}{m_Q^2 - k^2}G^{\mu \nu}(ux) \gamma_\nu \bigg]
    \end{split}
  \end{equation}
  where $\gamma_E$ is the Euler constant.
  
  For calculation of the correlator function(s) we need another ingredient of light-cone sum rules, namely the matrix elements of non-local operators $\bar{q}(x) \Gamma q(y)$ and $\bar{q}(x) \Gamma G_{\mu \nu} q(y)$ between vacuum and the K-meson , i.e. $\langle K(q) | \bar{q}(x) \Gamma q(y) | 0 \rangle$ and $\langle K(q) | \bar{q}(x) \Gamma G_{\mu\nu}q(y) | 0 \rangle$. Here $\Gamma$ is the any Dirac matrix, and $G_{\mu\nu}$ is the gluon field strength tensor, respectively. These matrix elements are defined in terms of K-meson distribution amplitudes (DA's). The DA's of K meson up to twist-$4$ are presented in \cite{Ball:2006wn}.
  
  From Eqs.~\eqref{eq:26} and \eqref{eq:27} it follows that the different Lorentz structures can be used for construction of the relevant sum rules. Among of six couplings, we need only $A_2 (A_2^{*})$ and $A_5 (A_5^{*})$ and $A_2 (A_2^{*})$ and $\tilde{A_5} (\tilde{A_5^{*}})$ for the cases a) and b) respectively. For determination of these coupling constants, we need to combine sum rules obtained from different Lorentz structures. From Eqs.\eqref{eq:26} and \eqref{eq:27} (for case a) it follows that the following Lorentz structures $\slashed{p} \slashed{q} \gamma_5$, $\slashed{p} \gamma_5$, $\slashed{q} \gamma_5$, $\gamma_5$, and $\slashed{p} \slashed{q} q_\mu$, $\slashed{p} q_ \mu$, $\slashed{q} q_\mu$, and $q_\mu$ appear. We denote the corresponding invariant functions $\Pi_1, \Pi_2, \Pi_3, \Pi_4$ and $\Pi_1^*, \Pi_2^*, \Pi_3^{*}$ and $\Pi_4^{*}$, respectively. Explicit expressions of the invariant functions $\Pi_i$ and $\Pi_i^*$ are very lengthy, and therefore we do not present them in the present study.
  
The sum rules for the corresponding strong coupling constants are obtained by choosing the coefficients aforementioned structures and equating to the corresponding results from hadronic and QCD sides. Performing doubly Borel transformation with respect to variable $p^2$ and $(p+q)^2$ in order to suppress the contributions of higher states and continuum we get the following four equations (for each transition).
  \begin{equation}
    \label{eq:24}
    \begin{split}
      \Pi_1^{B} &= -A_1^{(B)} - A_2^{(B)} + A_3^{(B)} + A_4^{(B)} -A_5^{(B)} +A_6^{(B)}, \\
      \Pi_2^{B} &= A_1^{(B)} (m_0 - m_0^\prime) + A_2^{(B)}(-m_1 - m_0^\prime) + A_3^{(B)} (-m_0 -m_{1}^\prime) + A_4^{(B)}(m_{1} - m_{1}^\prime) \\
      &~~~~~~+ A_5^{(B)}(-m_{2} - m_0^\prime) + A_6^{(B)} (m_{2} - m_{1}^\prime) \\
      \Pi_3^{B} &= A_1^{(B)} (-m_0^\prime) + A_2^{(B)}( -m_0^\prime) +A_3^{(B)} (-m_{1}^\prime) + A_4^{(B)}(-m_{1}^\prime) + A_5^{(B)}(-m_0^\prime) + A_6^{(B)}(-m_{1}^\prime) \\
      \Pi_4^{B} &= A_1^{(B)} (m_0 m_0^\prime - {m_0^\prime}^2) + A_2^{(B)}( -m_0^\prime m_{1} - {m_0^\prime}^2 )  + A_3^{(B)} (m_0 m_{1}^\prime + {m_{1}^\prime}^2)  \\
      &~~~~~~+ A_4^{(B)} (-m_{1}^\prime m_{1} + {m_{1}^\prime}^2) + A_5^{(B)}(-m_{0}^\prime m_2 - {m_{0}^\prime}^2) + A_6^{(B)}(-m_{2} m_{1}^\prime + {m_{1}^\prime}^2)
    \end{split}
  \end{equation}

  \begin{equation}
    \label{eq:25}
    \begin{split}
      \Pi_1^{*B} &= - \big\{ {A_1^*}^{(B)} + {A_2^*}^{(B)} - {A_3^*}^{(B)} -{A_4^*}^{(B)} + {A_5^*}^{(B)} - {A_6^*}^{(B)}\big\} , \\
      \Pi_2^{*B} &= -\big\{ {A_1^*}^{(B)} (m_0^* + {m_0}^\prime) + {A_2^*}^{(B)} ({m_0^\prime} - m_{1}^*) + {A_3^*}^{(B)} (-m_0^* + {m_{1}^\prime})  \\
      &~~~~~~+ {A_4^*}^{(B)} (m_0^* + {m_{1}^\prime}) + A_5^{(B)}(-{m_{2}}^* +{m_{0}^\prime}) + A_6^{(B)}({m_{1}^\prime} + {m_{2}}^* ) \big\}\\      
      \Pi_3^{*B} &= -\big\{ {A_1^*}^{(B)} {m_{0}^\prime}  + {A_2^*}^{(B)} {m_{0}^\prime} + {A_3^*}^{(B)} {m_{1}^\prime} + {A_4^*}^{(B)} {m_{1}^\prime} + {A_5^*}^{(B)} {m_{0}^\prime} + {A_6^*}^{(B)} {m_{1}^\prime} \big\}\\      
      \Pi_4^{*B} &= -\big\{ {A_1^*}^{(B)} (m_0^* {m_{0}^\prime} + {{m_{0}^\prime}}^2) + {A_2^*}^{(B)} ({{m_{0}^\prime}}^2 - {m_{1}^*} {m_{0}^\prime}) + {A_3^*}^{(B)} ({{-m_{1}^\prime}}^2 + {{m_{1}^\prime}} {{m_{0}^*}}) \\
      &~~~~~~ + {A_4^*}^{(B)} (-{{m_{1}^\prime}}^2 -{{m_{1}^\prime}} m_{1}^* )+ {A_5^*}^{(B)}({{m_{0}^\prime}}^2 - {m_{0}^\prime} {m_{2}^*}) + {A_6^*}^{(B)}(-{{m_{1}^\prime}}^2 - {m_{1}^\prime} {m_{2}^*} ) \big\}
    \end{split}
  \end{equation}%
  where superscript B means Borel transformed quantities,

  \begin{equation}
    \label{eq:22}
    A^{B(*)}_\alpha = g_{\alpha}^{(*)}  \lambda_{i} \lambda_{j}^\prime e^{-m_{i}^2/M_1^2 - m_{j}^2/M_2^2}.
  \end{equation}

  The masses of the initial and final baryons are close to each other, hence in the next discussions, we set $M_1^2 = M_2^2 = 2M^2$. In order to suppress the contributions of higher states and continuum we need subtraction procedure. It can be performed by using quark-hadron duality, i.e. starting some threshold the spectral density of continuum coincide with spectral density of perturbative contribution.  The continuum subtraction can be done using formula
  \begin{equation}
    \label{eq:12}
    (M^2)^n e^{- m_c^2/M^2} \rightarrow \frac{1}{\Gamma(n)} \int_{m^2}^{s_0} ds e^{-s/M^2} (s-m_c^2)^{n-1}
  \end{equation}
  For more details about continuum subtraction in light cone sum rules, we refer readers to work~\cite{Belyaev:1994zk}.
  
  As we have already noted in case a) we need to determine two coupling constants $g_2$($g_2^{*}$) and $g_5$($g_5^{*}$)  for each class of transitions. From Eqs \eqref{eq:24} and \eqref{eq:25} it follows that we have six unknown coupling constants but have only four equations. Two extra equations can be obtained by performing derivative over ($1/M^2$) of the any two equations. In result, we have six equations and six unknowns and the relevant coupling constants $g_2(g_2^*)$ and $g_5(g_5^{*})$  can be determined by solving this system of equations. 

  The results for scenario b) can be obtained from the results for scenario a) with the help of aforementioned replacements.
  
  From Eqs \eqref{eq:24} and \eqref{eq:25}, it follows that to estimate strong coupling constants $g_2(g_2^*)$ and $g_5(g_5^*)$  responsible for the decay of $\Omega_c \rightarrow \Xi_c K$ and $\Omega_c^{*} \rightarrow \Xi_c K$, we need the residues of  $\Omega_c$ and $\Xi_c$ baryons. For calculation of these residues for $\Omega_c$, we consider the following two point correlation functions

  \begin{equation}
    \label{eq:37}
    \begin{split}
      \Pi (p) &= \int d^4 x e^{ipx} \langle 0 | T \{\eta_{\Omega_c}(x) \bar{\eta}_{\Omega_c}(0) \} | 0 \rangle, \\ 
      \Pi^{\mu \nu} (p) &= \int d^4 x e^{ipx} \langle 0 | T \{\eta^{\mu}_{ \Omega_c}(x) \bar{\eta}^\nu_{\Omega_c}(0) \} | 0 \rangle . 
    \end{split}
  \end{equation}

  The interpolating currents $\eta_{\Omega_c}$ and $\eta_{\mu \Omega_c}$ couples not only to ground states, but also to negative (positive) parity excited states, therefore their contributions should be taken into account. In result, for physical parts of the correlation functions we get,
  \begin{equation}
    \label{eq:38}
    \begin{split}
      \Pi(p) & = \frac{\langle 0 | \eta | \Omega_{c}(p,s) \rangle \langle \Omega_{c}(p,s) | \bar{\eta}(0) | 0 \rangle}{-p^2 + m_{0}^2}
      + \frac{\langle 0 | \eta | \Omega_{1c}(p,s) \rangle \langle \Omega_{1c}(p,s) | \bar{\eta}(0) | 0 \rangle}{-p^2 + m_{1}^2}, \\
      & + \frac{\langle 0 | \eta | \Omega_{2(3)c}(p,s) \rangle \langle \Omega_{2(3)c}(p,s) | \bar{\eta}(0) | 0 \rangle}{-p^2 + m_{2}^2} + ...~,
    \end{split}
  \end{equation}

  \begin{equation}
    \label{eq:39}
    \begin{split}
      \Pi^{\mu \nu } &= \frac{\langle 0 | \eta_\mu | \Omega_c^{*}(p,s) \rangle \langle \Omega_c^{*}(p,s) | \bar{\eta}_\nu(0) | 0 \rangle}{m_{0}^{*^2} - p^2}
      + \frac{\langle 0 | \eta_\mu | \Omega_{1c}^{*}(p,s) \rangle \langle \Omega_{1c}^{*}(p,s) | \bar{\eta}_\nu(0) | 0 \rangle}{m_{1}^{*^2} - p^2} \\
      &+ \frac{\langle 0 | \eta_\mu | \Omega_{2(3)c}^{*}(p,s) \rangle \langle \Omega_{2(3)c}^{*}(p,s) | \bar{\eta}_\nu(0) | 0 \rangle}{m_{2}^{*2} - p^2} + ...
    \end{split}
  \end{equation}
where the dots denote contributions of higher states and continuum.
  The matrix elements in these expressions are defined as
  \begin{equation}
    \label{eq:40}
    \begin{split}
      \langle 0 | \eta | \Omega_c(p,s) \rangle & = \lambda_{0} u(p), \\
      \langle 0 | \eta | \Omega_{1(2)c}(p,s) \rangle & = \lambda_{1(2)} \gamma_5 u(p) \\
      \langle 0 | \eta | \Omega_c^{*}(p,s) \rangle & = \lambda_{0}^{*} u_\mu(p), \\
      \langle 0 | \eta | \Omega_{1(2)c}(p,s) \rangle & = \lambda_{1(2)}^{*} \gamma_5 u_\mu(p) . 
    \end{split}
  \end{equation}


  As we already noted, only the structure $g_{\mu \nu}$ describes the contribution coming from $3/2$ baryons. Therefore we retain only this structure.
  
For the physical parts of the correlation function, we get
  \begin{equation}
    \label{eq:42}
    \begin{split}
      \Pi^{phy} &= \frac{(\slashed{p} + m_0) \lambda_0^2 }{m_0^2 - p^2} + \frac{(\slashed{p} - m_{1}) \lambda_{1}^2 }{m_{1}^2 - p^2} + 
      \frac{(\slashed{p} \mp m_{2(3)} )\lambda_{2(3)}^{2} }{m_{2(3)}^{2} - p^2} \\
      \Pi^{phy}_{\mu \nu} &= \frac{(\slashed{p} + m_0^*) g_{\mu \nu} \lambda_0^{*^2} }{m_0^{*^2} - p^2} + \frac{(\slashed{p} - m^*_{1}) g_{\mu \nu} \lambda_{1}^{*^2} }{m_{1}^{*^2} - p^2} +
      \frac{(\slashed{p} \mp m_{2(3)}^{*}) \lambda_{2(3)}^{2} }{m_{2(3)}^{*^2} - p^2}. 
    \end{split}
  \end{equation}

  Here in the last term, upper(lower) sign corresponds to case a) (case b).

  Denoting the coefficients of the Lorentz structures $\slashed{p}$ and $I$ operators $\Pi_1$, $\Pi_2$ and $\slashed{p} g_{\mu \nu}$, $g_{\mu \nu}$ as $\Pi_1^*$, $\Pi_2^*$ respectively and performing Borel transformations with respect to $- p^2$, for spin $1/2$ case we find,

  \begin{equation}
    \label{eq:43}
    \begin{split}
      \Pi_1^{B} &=  \lambda_{0}^2 e^{-m_{0}^2/M^2} + \lambda_{1}^2 e^{-m_{1}^2/M^2} + \lambda_{2(3)}^{2} e^{-m_{2}^{2}/M^2}, \\
      \Pi_2^{B} &=  \lambda_{0}^2 m_{0}e^{-m_{0}^2/M^2} - \lambda_{1}^2 m_{1} e^{-m_{1}^2/M^2} \mp \lambda_{2(3)}^{2} m_{2(3)} e^{-m_{2(3)}^{ 2}/M^2}. 
    \end{split}
  \end{equation}

  The expressions for spin $3/2$ case formally can be obtained from these expressions by replacing $\lambda \rightarrow \lambda^{*}$, $m \rightarrow m^{*}$ and $\Pi \rightarrow \Pi^*$.
  The invariant functions $\Pi_i$, $\Pi_i^*$ from QCD side can be calculated straightforwardly by using the operator product expansion. Their expressions are presented in~\cite{Aliev:2009jt} (see also \cite{Aliev:2017led}).

 Similar to the determination of the strong coupling constant, for obtaining the sum rule for residues we need the continuum subtraction. It can be performed in following way. In terms of the spectral density $\rho(s)$ the Borel transformed $\Pi^B$ can be written as
  \begin{equation}
    \label{eq:14}
    \begin{split}
      \Pi_i^B = \int_{m_c^2}^{\infty} \rho_i(s) e^{-s / M^2} ds 
    \end{split}
  \end{equation}
  The continuum subtraction can be done by using the quark-hadron duality and for this aim it is enough to replace
  \begin{equation}
    \label{eq:15}
    \int_{m_c^2}^{\infty} \rho_i(s) e^{-s / M^2} ds \rightarrow \int_{m_c^2}^{s_0} \rho_i(s) e^{-s / M^2} ds. 
  \end{equation}
  
  It follows from the sum rules we have only two equations, but six (three masses and three residues) unknowns. In order to simplify the calculations, we take the masses of $\Omega_c$ as input parameters. Hence, in this situation, we need only one extra equation, which can be obtained by performing derivatives over ($\frac{-1}{M^2}$) on both sides of the equation. Note that the residues of $\Xi_c$ baryons are calculated in a similar way.
  \section{Numerical Analysis}
\label{sec:numeric}
In this section we present our numerical results of the sum rules for the strong coupling constants responsible for $\Omega_c(3000) \rightarrow \Xi_c^{+} K^-$ and $\Omega_c(3066) \rightarrow \Xi_c^{+} K^-$ decays derived in previous section. The Kaon distribution amplitudes are the key non-perturbative inputs of sum rules whose expressions are presented in \cite{Ball:2006wn}. The values of other input parameters are:
\begin{equation}
  \label{eq:7}
  \begin{split}
    f_K & = 0.16~\rm{GeV}, \\
    m_0^2 & = (0.8 \pm 0.2) ~\rm{GeV^2}, \\
    \langle \bar{q} q \rangle & = -(0.240 \pm 0.001)^3~\rm{GeV^3}, \\
    \langle \bar{s}s \rangle & = 0.8~ \langle \bar{q} q \rangle . 
  \end{split}
\end{equation}

\begin{figure}[hbt]
  \centering
  \includegraphics[scale=0.7]{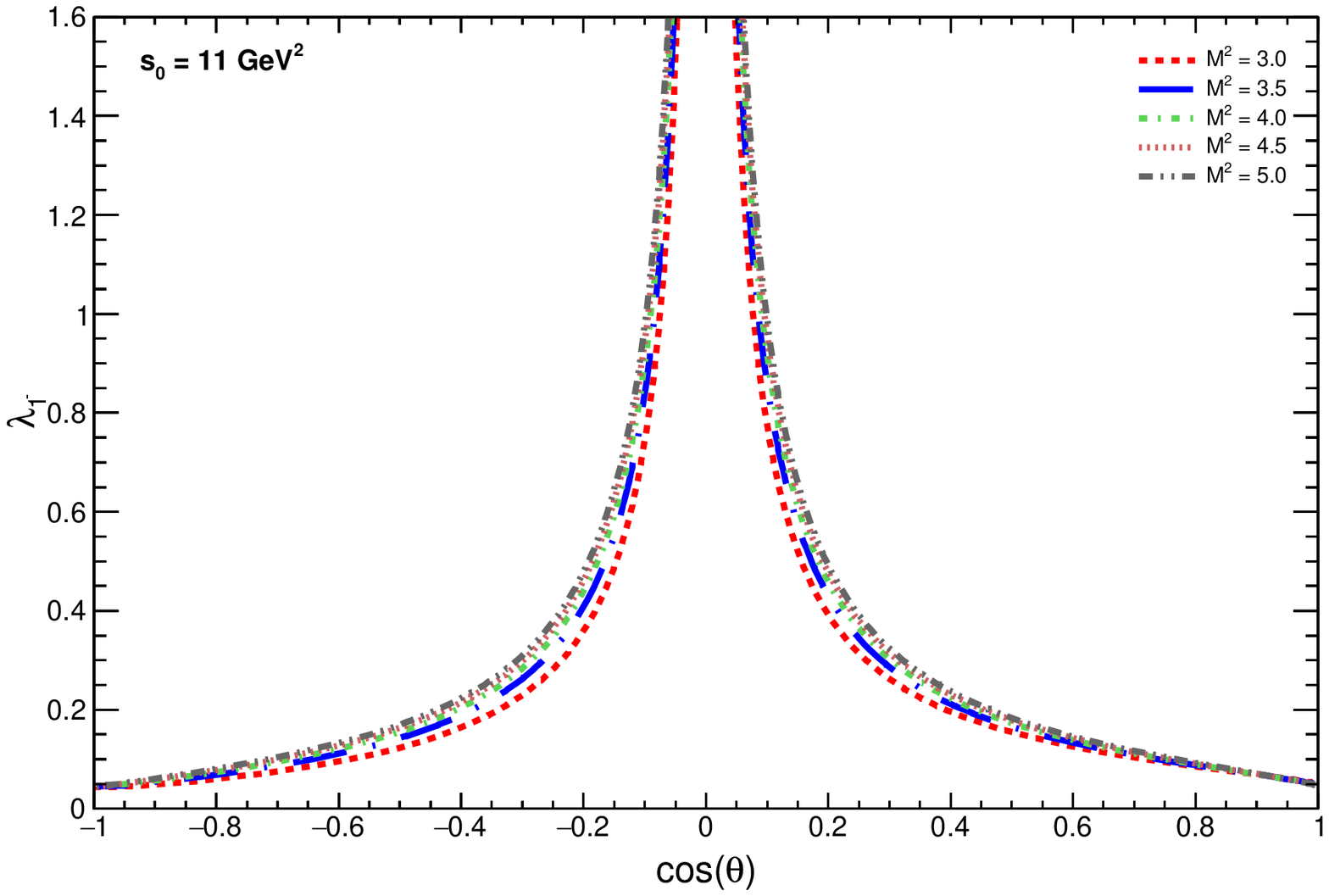}
  \caption{The dependence of residue for $\Omega_c(3000)$ on $\cos{\theta}$ at $s_0 = 11~\rm{GeV^2}$ and at various fixed values of $M^2$ for scenario a).}
  \label{fig:1}
\end{figure}

\begin{figure}[hbt]
  \centering
  \includegraphics[scale=0.7]{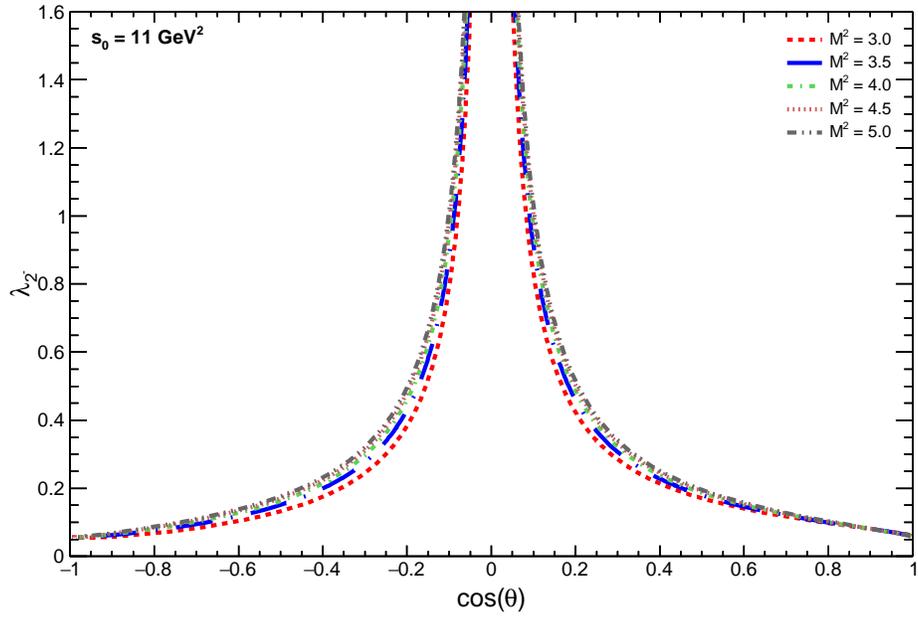}
  \caption{Same as in Fig.~\ref{fig:1}, but for $\Omega_c(3050)$.}
  \label{fig:2}
\end{figure}

The sum rules for $g_{-+}$ and $g_{-+}^{*}$contain the continuum threshold $s_0$, Borel variable $M^2$ and parameter $\beta$ in interpolating current for spin $1/2$ particles. In order to extract reliable values of these constants from QCD sum rules, we must find the working regions of $s_0,~M^2$ and $\beta$ in such a way that the result is insensitive to the variation of these parameters. The working region of $M^2$ is determined from conditions that the operator product expansion (OPE) series be convergent and higher states and continuum contributions should be suppressed. More accurately, the lower bound of $M^2$ is obtained by demanding the convergence of OPE and dominance of the perturbative contributions over the non-perturbative one. The upper bound of $M^2$ is determined from the condition that the pole contribution should be larger than the continuum and higher states contributions. We obtained that both conditions are satisfied when $M^2$ lies in the range
\begin{equation}
  \label{eq:8}
  2.5~\rm{GeV^2} \leq M^2 \leq 5~\rm{GeV^2}.
\end{equation}
The continuum threshold $s_0$ is not arbitrary and related with the energy of the first excited state i.e. $s_0 = (m_{\text{ground}} + \delta)^2$. Analysis of various sum rules shows that $\delta$ varies between $0.3$ and $0.8$ \rm{GeV}, and in this analysis $\delta=0.4~\rm{GeV}$ is chosen.
As an example, in Figs.~\ref{fig:1} and \ref{fig:2} we present the dependence of the residues of $\Omega_c(3000)$ and $\Omega_c(3050)$ on $\cos{\theta}$ for the scenario a)  at $s = 11~\rm{GeV^2}$ and several fixed values of $M^2$ , respectively. From these figures, we obtain that when $\cos{\theta}$ lies between $-1$ and $-0.5$ the residues exhibits good stability with respect to the variation of $\cos{\theta}$ and the results are practically insensitive to the variation of $M^2$. And we deduce the following results for the residues
\begin{equation}
  \label{eq:44}
  \begin{split}
    \lambda_{1} &= (0.08 \pm 0.03)~~\rm{GeV^3},  \\
    \lambda_{2} &= (0.11 \pm 0.04)~~\rm{GeV^3}.
  \end{split}
\end{equation}

\begin{figure}[hbt]
  \centering
  \includegraphics[scale=0.7]{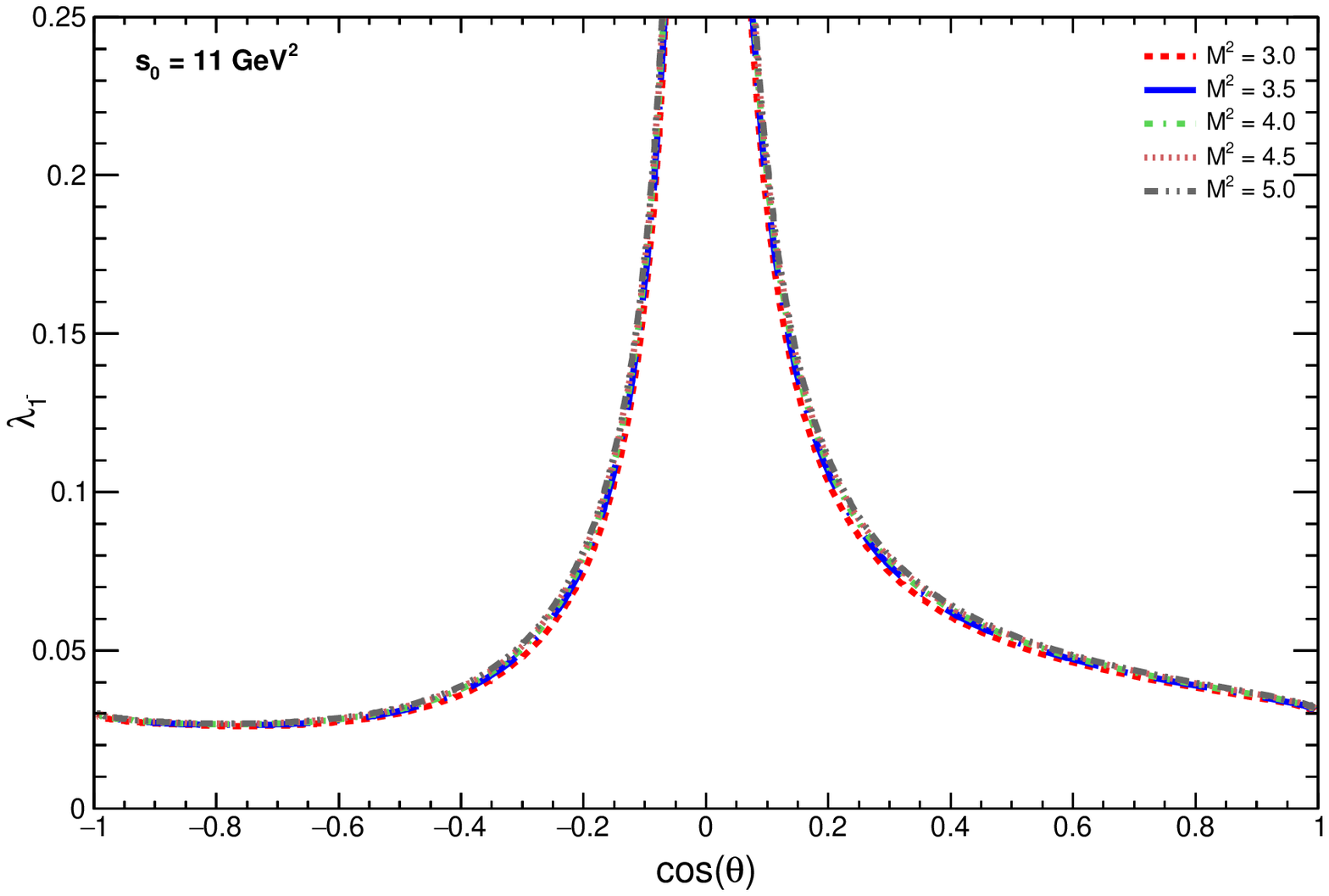}
  \caption{Same as in Fig.~\ref{fig:1}, but for scenario b).}
  \label{fig:3}
\end{figure}

\begin{figure}[hbt]
  \centering
  \includegraphics[scale=0.7]{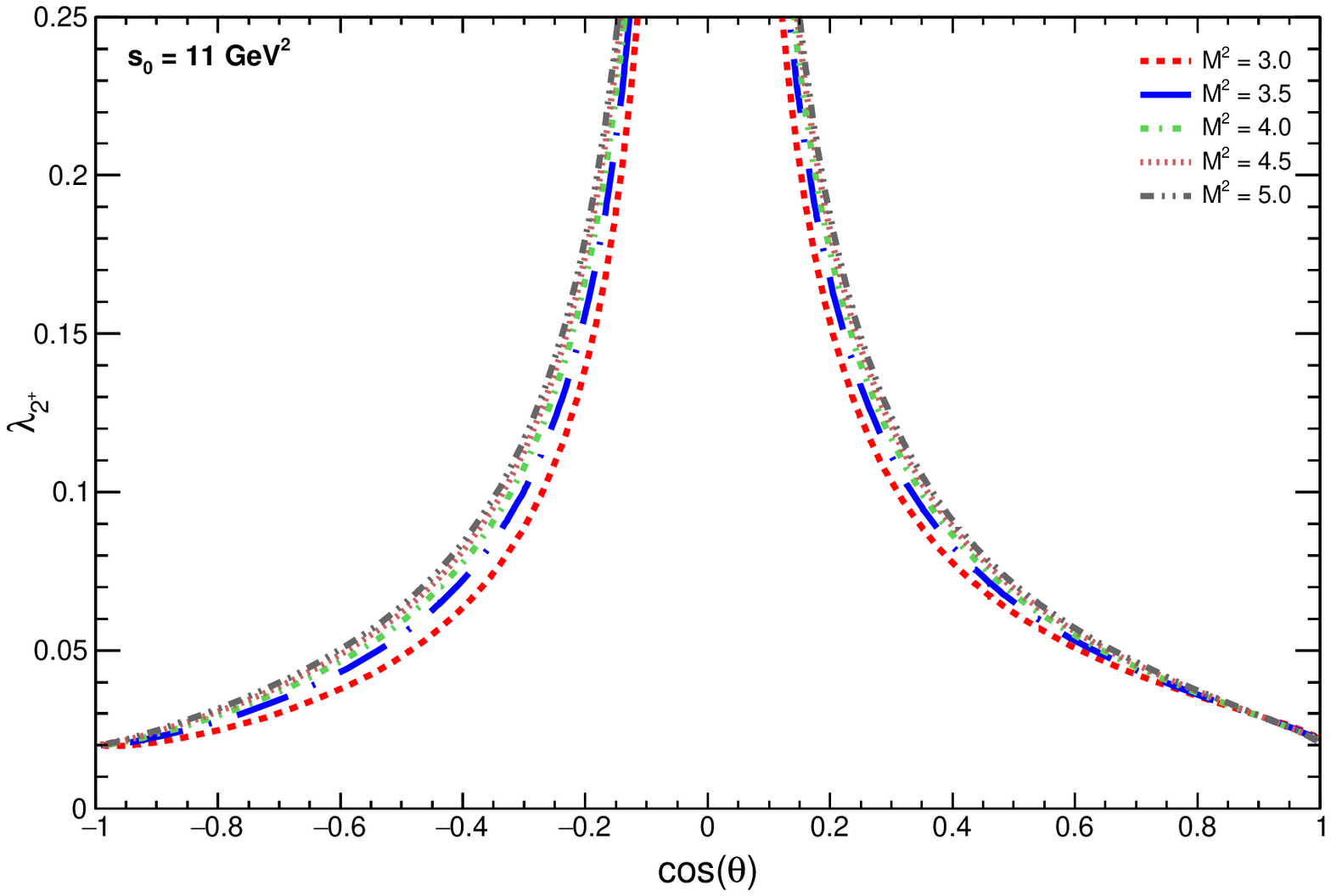}
  \caption{Same as in Fig.~\ref{fig:2}, but for scenario b).}
  \label{fig:4}
\end{figure}

Performing similar analysis for $\Omega_c$ baryons in scenario b) we get (Figs.~\ref{fig:3} and \ref{fig:4})

\begin{equation}
  \label{eq:45}
  \begin{split}
     \lambda_{1} &= (0.030 \pm 0.001)~~\rm{GeV^3}, \\
    \lambda_{3} &= (0.04 \pm 0.01)~~\rm{GeV^3}. 
  \end{split}
\end{equation}
The detailed numerical calculations lead to the following results for spin $3/2$ $\Omega_c$ baryon residues:

\begin{equation}
  \label{eq:46}
  \begin{split}
    \lambda_{1}^* &= (0.18 \pm 0.02)~~\rm{GeV^3}, \\
    \lambda_{2}^* &= (0.17 \pm 0.02)~~\rm{GeV^3}, \\
  \end{split}
\end{equation}

\begin{equation}
  \label{eq:47}
  \begin{split}
    \lambda_{1}^* &= (0.024 \pm 0.002)~~\rm{GeV^3}, \\
    \lambda_{3}^* &= (0.05 \pm 0.01)~~\rm{GeV^3}. 
  \end{split}
\end{equation}

From these results we observe that the residues of $\Omega_c$ baryons in scenario a) is larger than that one for the scenario b). This leads to the larger strong coupling constants for scenario b) because it is inversely proportional to the residue.

\begin{figure}[hbt]
  \centering
  \includegraphics[scale=0.7]{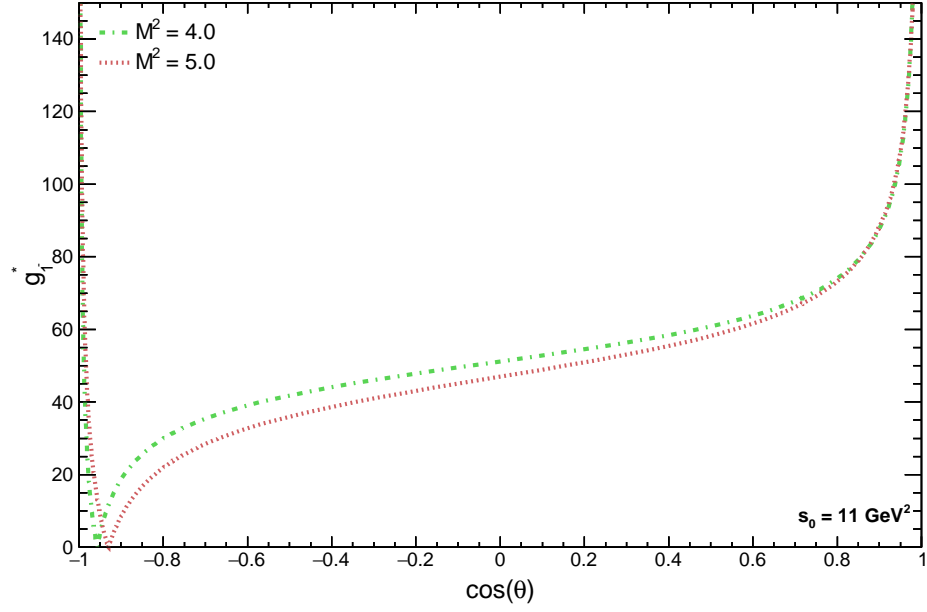}
  \caption{The dependence of strong coupling constant for $\Omega_c(3066) \rightarrow \Xi_c K$ on $\cos{\theta}$ at $s_0 = 11~\rm{GeV^2}$ and at three fixed values of $M^2$ for scenario a).}
  \label{fig:5}
\end{figure}

\begin{figure}[hbt]
  \centering
  \includegraphics[scale=0.7]{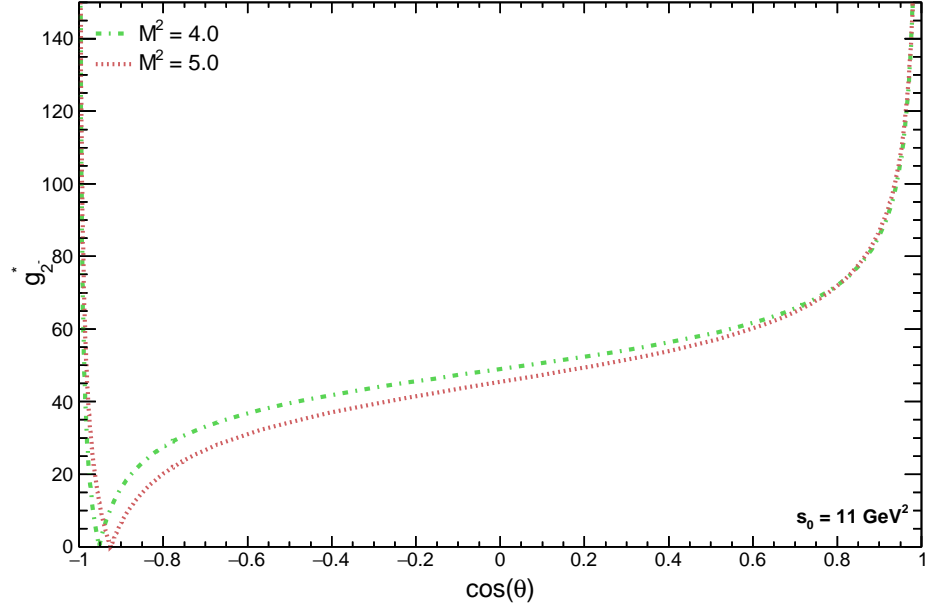}
  \caption{Same as in Fig.~\ref{fig:5}, but for $\Omega_c(3090) \rightarrow \Xi_c K$ transition.}
  \label{fig:6}
\end{figure}

\begin{figure}[hbt]
  \centering
  \includegraphics[scale=0.7]{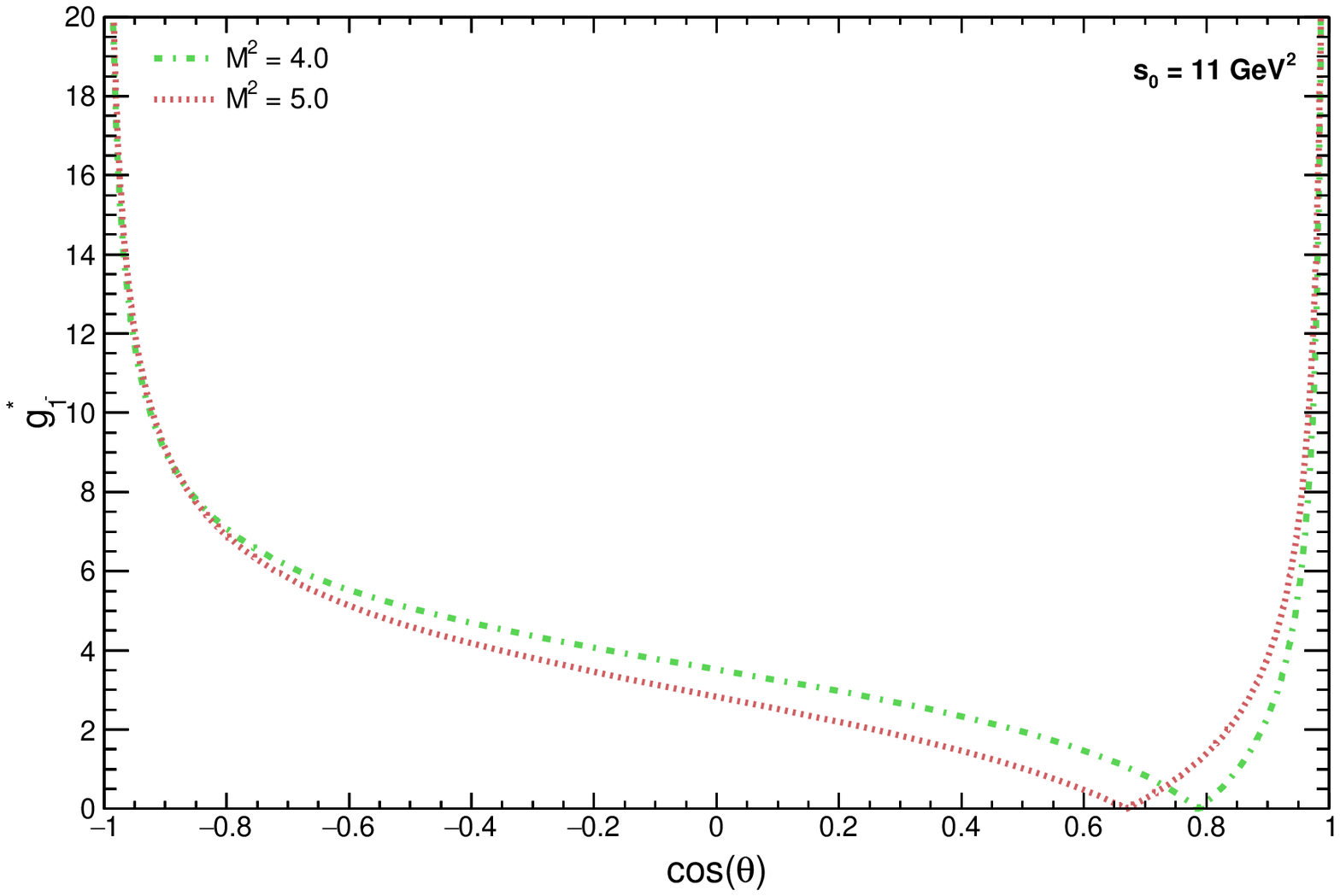}
  \caption{Same as in Fig.~\ref{fig:5}, but for scenario b).}
  \label{fig:7}
\end{figure}

\begin{figure}[hbt]
  \centering
  \includegraphics[scale=0.7]{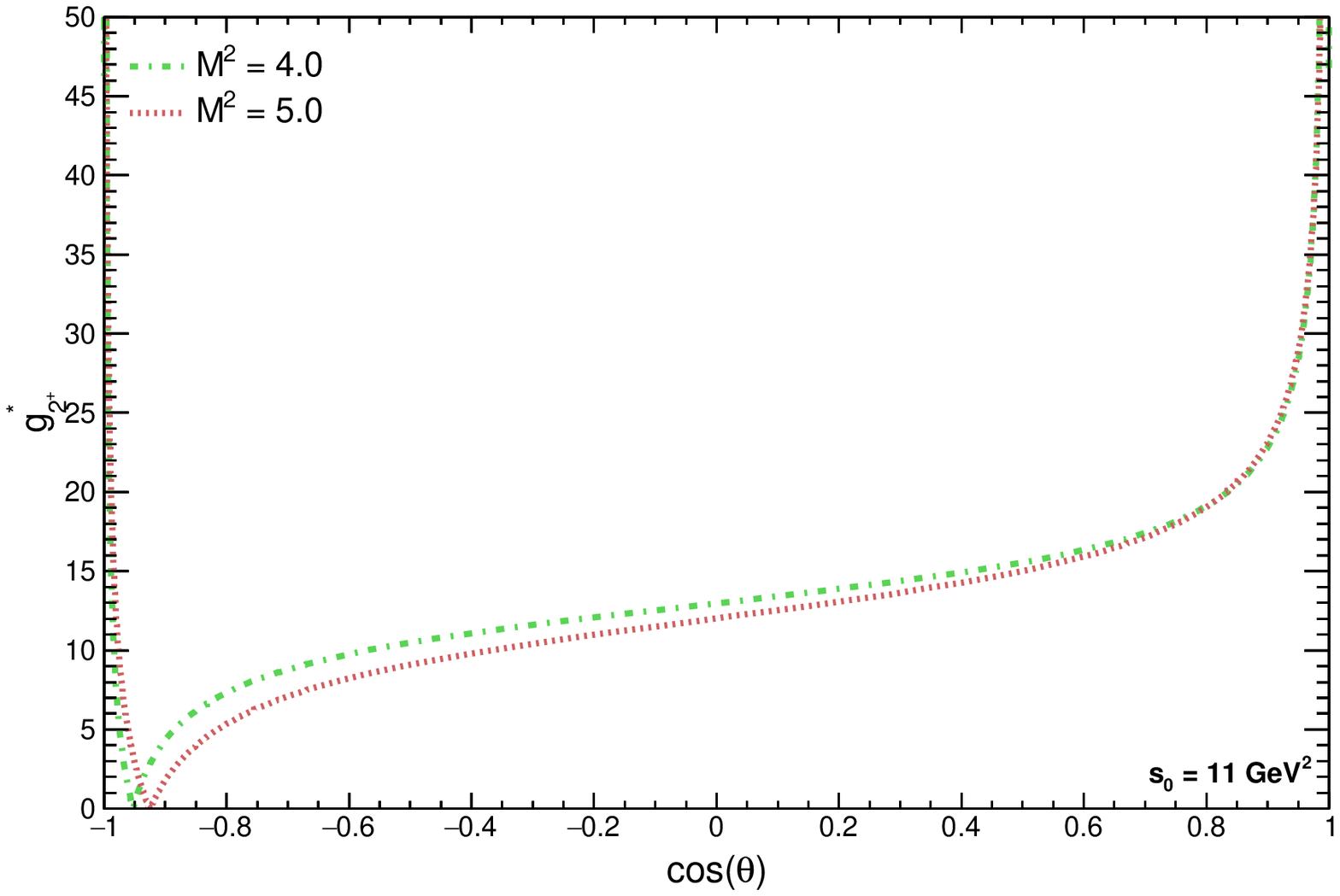}
  \caption{Same as in Fig.~\ref{fig:6}, but for scenario b).}
  \label{fig:8}
\end{figure}

Having obtained the values of the residues, our next problem is the determination of the corresponding coupling constants using the values of $M^2$ and $s_0$ in their respective working regions which are determined from mass sum rules . In Figs.~\ref{fig:5},~\ref{fig:6},~\ref{fig:7}, and \ref{fig:8}, we studied the dependence of the strong coupling constants for $\Omega_c^{*} \rightarrow \Xi_c K^0$ transitions for the scenarios a) and b) on $\cos{\theta}$, respectively. We obtained that when $M^2$ varies in its working region the strong coupling constant demonstrates weak dependence on $M^2$, and the results for the spin-$3/2$ states also practically do not change with the variation of $s_0$. Our results on the coupling constants are:

For scenario a:
\begin{equation}
  \label{eq:46}
  \begin{split}
    g_{2} & = 19 \pm 2  \hspace{2cm}  g_{2}^*  = 40 \pm 10 \\ 
    g_{5} & = 20 \pm 2 \hspace{2cm}  g_{5}^*  = 42 \pm 10 
  \end{split}
\end{equation}

For scenario b:
\begin{equation}
  \label{eq:47}
  \begin{split}
    g_{2} & = 2.2 \pm 0.2  \hspace{2cm}  g_{2}^*  = 2.0 \pm 0.5 \\ 
    \tilde{g_5} & = 6 \pm 1 \hspace{2.65cm}  \tilde{g_5}^* = 8 \pm 1 
  \end{split}
\end{equation}

The decay widths of these transitions can be calculated straightforwardly and we we get;
\begin{equation}
  \label{eq:48}
  \begin{split}
  \Gamma &= \frac{g_i^2}{16 \pi m_i^3} \big[ (m_i + m_{0}^ \prime)^2 - m_K^2 \big] \lambda^{1/2}(m_{i}^2, m_{0}^{\prime ^2}, m_K^2) \\
  \Gamma &= \frac{g_i^{*2}}{192 \pi m_i^{*^5}} \big[ (m_i^* + m_{0}^\prime)^2 - m_K^2 \big] \lambda^{3/2}(m_{i}^{*^2}, m_{0}^{ \prime ^2}, m_K^2) 
  \end{split}
  \end{equation}
  where $m_i (m_i^*)$ and $m_0^\prime$ are the mass of initial spin $1/2$ (spin $3/2$) $\Omega_c$ baryon and $\Xi_c$ baryons respectively and $\lambda(x,y,z) = x^2 + y^2 + z^2 - 2 x y - 2 x z - 2 y z$.
  Having the relevant strong coupling constants, the  decay width values for scenario a) and b) are shown in Table~\ref{tab:1}.
\begin{table}
  \begin{tabular}{ccc}
    \toprule
    \multirow{2}{*}{} & Scenario-a & Scenario-b  \\
                       & (GeV) & (GeV) \\
    \midrule
     $ \Gamma (\Omega_c(3000) \rightarrow \Xi_c K)$ & $8.1 \pm 1.8 $ & $0.10 \pm 0.02$ \\
     $ \Gamma (\Omega_c(3050) \rightarrow \Xi_c K)$ & $14.1 \pm 3$ & $(3.8 \pm 1.2) \times 10^{-3}$ \\
     $ \Gamma (\Omega_c(3066) \rightarrow \Xi_c K)$ & $(6.6 \pm 3) \times 10^{-3}$ & $(1.6 \pm 0.6) \times 10^{-5}$\\
     $\Gamma (\Omega_c(3090) \rightarrow \Xi_c K)$ & $(1.3 \pm 0.5) \times 10^{-2}$ & $0.10 \pm 0.04$ \\
    \bottomrule
  \end{tabular}
  \caption{Decay widths for the two-scenarios considered are shown.}
      \label{tab:1}
\end{table}

 Our results on the decay widths are also drastically different than the one presented in \cite{Agaev:2017lip}. In our opinion, the source of these discrepancies are due to the following facts:
\begin{itemize}
\item In \cite{Agaev:2017lip}, the contributions coming from $\Xi_c^-$ baryons are all neglected.
\item The second reason is due to the procedure presented in \cite{Agaev:2017lip}, namely by choosing the relevant threshold $s_0$, isolating the contributions of the corresponding $\Omega_c$ baryons is incorrect. From analysis of various sum rules, it follows that $s_0 = (m_{\text{ground}} + \delta)^2$, where $0.3~\rm{GeV} \leq \delta \leq 0.8~\rm{GeV}$. Since the mass difference between $\Omega_c(3000)$ and $\Omega_c(3090)$ is around $0.1~\rm{GeV}$, isolating the contribution of each baryon is impossible while their contributions should be taken into account simultaneously. For these reasons our results on decay widths are different than those one predicted in \cite{Agaev:2017lip}. From experimental data on the width of $\Omega_c$ are~\cite{Olive:2016xmw}:
\begin{equation}
  \label{eq:9}
  \begin{split}
      \Gamma (\Omega_c(3000) \rightarrow \Xi_c^+ K^-) &= (4.5 \pm 0.6 \pm 0.3)~\rm{MeV} \\
      \Gamma (\Omega_c(3050) \rightarrow \Xi_c^+ K^-) &= (0.8 \pm 0.2 \pm 0.1)~\rm{MeV} \\
      \Gamma (\Omega_c(3066) \rightarrow \Xi_c^+ K^-) &= (3.5 \pm 0.4 \pm 0.2)~\rm{MeV} \\
     \Gamma (\Omega_c(3090) \rightarrow \Xi_c^+ K^-) &= (8.7 \pm 1.0 \pm 0.8)~\rm{MeV} \\
     \Gamma (\Omega_c(3119) \rightarrow \Xi_c^+ K^-) &= (1.1 \pm 0.8 \pm 0.4)~\rm{MeV} 
  \end{split}
\end{equation}
We find out that, our predictions strongly differ from the experimental results.

By comparing our predictions with the experimental data, we conclude that both scenarios are ruled out.
\end{itemize}

\section{Conclusion}
\label{sec:conclusion}

In conclusion, we calculated the strong coupling constants of negative parity $\Omega_c$ baryon with spins $1/2$ and $3/2$ with $\Xi_c$ and $K$ meson in the framework of light cone QCD sum rules. Using the obtained results on coupling constants we estimate the corresponding decay widths. We find that our predictions on the decay widths under considered scenarios are considerably different from experimental data as well as theoretical predictions and considered both scenarios are ruled out. Therefore further theoretical studies for determination of the quantum numbers of $\Omega_c$ states as well as for correctly reproducing the decay widths of $\Omega_c$ baryons are needed.

\section{Acknowledgments}
The authors acknowledge METU-BAP Grant no. 01-05-2017004.

\bibliography{QCD}

\end{document}